# Inflammation-driven glial alterations in the cuprizone mouse model probed with diffusion-weighted magnetic resonance spectroscopy at 11.7 T

Guglielmo Genovese[a,b], Marco Palombo[c], Mathieu D. Santin[a,b], Julien Valette[d,e], Clémence Ligneul[f], Marie-Stéphane Aigrot[b,g], Nasteho Abdoulkader[a], Dominique Langui[b,g], Aymeric Millecamps[g], Anne Baron-Van Evercooren[b], Bruno Stankoff[b], Stéphane Lehericy[a,b], Alexandra Petiet*[a,b] and Francesca Branzoli*[a,b]

[a] Paris Brain Institute (Institut du Cerveau – ICM), Center for Neuroimaging Research - CENIR, F-75013, Paris, France

[b] Hopital Pitié-Salpêtrière, ICM, Sorbonne Université, Inserm U 1127, CNRS UMR 7225, F-75013, Paris, France

[c] Centre for Medical Image Computing and Department of Computer Science, University College London, London, United Kingdom

[d] Commissariat à l'Energie Atomique et aux Energies Alternatives (CEA), MIRCen, F-92260, Fontenay-aux-Roses, France

[e] Neurodegenerative Diseases Laboratory, UMR9199, CEA, CNRS, Université Paris Sud, Université Paris-Saclay, F-92260, Fontenay-aux-Roses, France

[f] Wellcome Centre for Integrative Neuroimaging, FMRIB, Nuffield Department of Clinical Neurosciences, University of Oxford, Oxford, United Kingdom

[g] Core Facility ICM.Quant, Institut du Cerveau - ICM, Paris, France.

*The authors equally contributed to the study

**Corresponding author:**

Francesca Branzoli, Ph.D.
Institut du cerveau - ICM
Hôpital Pitié-Salpetrière
47 boulevard de l'Hôpital
CS 21414
75646 Paris Cedex 13
Phone number: +33 (0)1 57 27 46 46
Email: francesca.branzoli@icm-institute.org

**Abstract**

Inflammation of brain tissue is a complex response of the immune system to the presence of toxic compounds or to cell injury, leading to a cascade of pathological processes that include glial cell activation. Noninvasive magnetic resonance imaging markers of glial reactivity would be very useful for *in vivo* detection and monitoring of inflammation processes in the brain, as well as for evaluating the efficacy of personalized treatments. Due to their specific location in glial cells, *myo*-inositol (mIns) and choline compounds (tCho) seem the best candidates for probing glial-specific intra-cellular compartments. However, their concentrations quantified using conventional proton magnetic resonance spectroscopy (MRS) are not specific for inflammation. In contrast, it has been recently suggested that mIns intra-cellular diffusion, measured using diffusion-weighted MRS (DW-MRS) in a mouse model of reactive astrocytes, could be a specific marker of astrocytic hypertrophy. In order to evaluate the specificity of both mIns and tCho diffusion to inflammation-driven glial alterations, we performed DW-MRS in the corpus callosum of cuprizone-fed mice after 6 weeks of intoxication and evaluated the extent of astrocytic and microglial alterations using immunohistochemistry. Both mIns and tCho apparent diffusion coefficients were significantly elevated in cuprizone-fed mice compared to control mice, while histologic evaluation confirmed the presence of severe inflammation. Additionally, mIns and tCho diffusion showed, respectively, strong and moderate correlations with histological measures of astrocytic and microglial area fractions, confirming DW-MRS as a promising tool for specific detection of glial changes under pathological conditions.

# 1. Introduction

Inflammation is the immune system's response to various pathological conditions such as the presence of toxic compounds, damaged cells, or pathogens, and it is a common denominator of several brain diseases. Brain inflammatory responses involve a cascade of complex mechanisms including glial cell activation and it has been suggested to have both detrimental and beneficial effects (Martino et al., 2002; Sochocka et al., 2017). While acute inflammatory processes are thought to minimize ongoing injury or infection and facilitate restoration of tissue homeostasis, chronic inflammation may lead to secondary injury and contribute to neurodegeneration (González and Pacheco, 2014; Khandelwal et al., 2011).

Detecting inflammation processes noninvasively would be highly beneficial for planning optimal therapeutic strategies and for monitoring the effect of anti-inflammatory treatments in several brain pathologies. However, specific, noninvasive markers of glial cell response to injury are still lacking. While diffusion MRI methods mostly target neuro-axonal integrity (Alexander et al., 2019; Jelescu and Budde, 2017), proton magnetic resonance spectroscopy ($^1$H MRS) enables the specific study of different cell populations, including glial cells, thanks to the cell-type specific compartmentalization of brain metabolites. In particular, due to the preferential location of choline compounds (tCho) and myo-inositol (mIns) in glial cells, increased levels of tCho and mIns have been linked to inflammation and gliosis, respectively (Öz et al., 2014). However, changes in tCho and mIns concentrations could be related to concomitant pathological mechanisms, involving variations in glial cell density and cellular metabolic dysfunction, and, to our knowledge, they have never been proved as specific markers of inflammation processes. $^1$H MRS has been used to monitor longitudinal glial and axonal changes in the corpus callosum of cuprizone (CPZ)-fed mice (Orije et al., 2015). In the CPZ mouse model, activated astrocytes and microglia initiate inflammatory processes within the first 2-3 weeks of intoxication, in response to apoptosis of oligodendrocytes and onset of demyelination (Gudi et al., 2014; Matsushima and Morell, 2006; Remington et al., 2007). After six weeks of cuprizone treatment, mice usually exhibit severe demyelination associated with a significant inflammatory response (Morell et al., 1998). In particular, it has been suggested that the corpus callosum is the region most affected by the CPZ diet (Goldberg et al., 2015; Gudi et al., 2009). Biochemical longitudinal changes detected using MRS in the corpus callosum of CPZ mice have been attributed to reversible demyelination, axonal injury, and inflammation, without specificity for either lesion (Orije et al., 2015). Interestingly, tCho levels were reduced in CPZ mice after 6 weeks of intoxication compared to healthy mice and mIns levels were unchanged, while histology proved evidence of ongoing inflammation.

In contrast to conventional MRS, diffusion-weighted MRS (DW-MRS) informs on cell-specific microstructural abnormalities by probing the motion of several metabolites in brain tissue (Nicolay et al., 2001; Palombo et al., 2018; Ronen and Valette, 2015). DW-MRS has been suggested to be more sensitive to glial reactivity than classical MRS from both clinical and preclinical investigations. Specifically, tCho and mIns diffusion alterations have been linked to glial cell hypertrophy driven by inflammation process (Ercan et al., 2016; Ligneul et al., 2019). On the other hand, the diffusion of the neuronal metabolite *N*-acetylaspartate, co-measured with *N*-acetylaspartylglutamate, (tNAA) has been proposed as a marker of intra-axonal integrity (Wood et al., 2012).

The main goals of this study were to validate DW-MRS markers of inflammation with histological markers of glial cells and to compare their utility with respect to conventional MRS measurements. To this aim, we performed single-voxel DW-MRS experiments in the corpus callosum of control and CPZ-fed mice at 11.7 T and correlated apparent diffusion coefficients (ADCs) and concentrations of tCho and mIns with histological markers of microglia and astrocytes derived from immunohistochemistry (IHC). The CPZ mouse model is a very well-characterized model of demyelination, which is associated with severe inflammation (Kipp et al., 2009; Skripuletz et al., 2011). Our working hypotheses are illustrated in Figure 1. We expected that activated astrocytes and microglia would be reflected in significantly higher mIns and tCho diffusion in CPZ mice after 6 weeks of intoxication compared to control mice, and that these changes would be accompanied by significant alterations in glial cell markers measured with IHC. In addition, based on previous DW-MRS findings (Ligneul et al., 2019), we hypothesized that only mIns would be massively present in astrocytes. We also hypothesized that the diffusivity of tNAA would be unchanged due to intact intra-axonal compartments. In contrast, both total creatine – creatine (Cr) + phosphocreatine (PCr) – (tCr) diffusion and metabolite concentrations were not expected to be specific markers of any pathological process in the CPZ mouse model, due to the concurrent presence of other alterations related to demyelination, such as oligodendrocyte death and mitochondrial dysfunction, which may complicate the interpretation of possible changes of these markers. Our results confirmed DW-MRS as a promising method to detect inflammatory-driven glial changes in pathology.

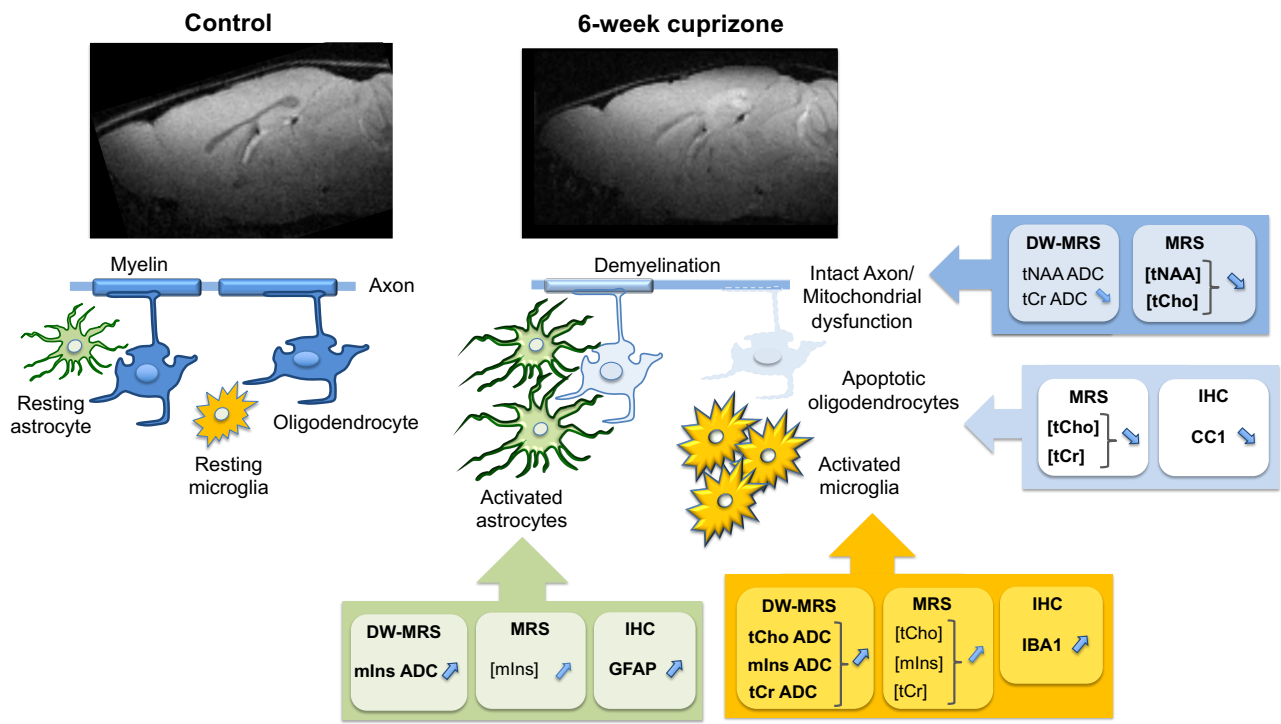

**Figure 1: Working hypotheses.** Schematic drawing of the known cellular alterations in CPZ mice after 6-week intoxication and expected corresponding changes in MRS and DW-MRS markers (right) compared to control mice (left). Stronger hypotheses are highlighted with bold characters. ADC: apparent diffusion coefficient; CC1: adenomatous polyposis coli oligodendrocyte proteins; GFAP: glial fibrillary acidic protein; IBA1: adaptive molecule linking ionizing calcium 1; mIns: *myo*-inositol; tCho: choline compounds; tCr : creatine + phosphocreatine ; tNAA: *N*-acetylaspartate + *N*-acetylaspartylglutamate.

## 2. Material and methods

**2.1 Animal experiments**

All animal experiments were performed in accordance with the EU Directive 2010/63/EU for animal experiments and approved by the Charles Darwin's Ethical Committee for Animal Experimentation (#6119-2016071314309890).

Eighteen 8-week old C57Bl/6J female mice were included in the study. Female mice were chosen because they are easier to breed compared to males, whereas CPZ-induced cellular changes are similar in males and females (Taylor et al., 2010). Eight mice were fed with 0.2% CPZ for 6 weeks, while 10 mice were fed with a normal chow diet and used as healthy control mice. All mice were fed *ad libidum* (Pfeifenbring et al., 2015). Animal care was provided by a veterinarian and animal technicians.

This study was part of a larger multimodal MRI study (MAXIMS: Myelin and AXon Imaging Markers Specificity) aiming at validating, through direct comparison with histology, noninvasive imaging markers of inflammation and axonal degeneration in the CPZ model. The mice were therefore split in two groups. After the imaging session, half of them were sacrificed for electron microscopy (EM) assessment and the other half for IHC assessment. The MRI markers were correlated with EM markers, and the results were the object of an independent article (Hill et al., 2019). Here, we present the results obtained from DW-MRS experiments, which were correlated with astrocyte- and microglia-specific IHC markers.

**2.2 Experimental procedures**

Prior to the MRI/MRS experiments, the mice were anesthetized with 4% isoflurane in a mixture of air (flow rate: 1 l/min) and oxygen (flow rate: 0.2 l/min) for induction. The mice were then placed on a stereotaxic bed with a bite bar and two ear bars. During the experiments, the level of anesthesia was maintained between 1 and 1.5% to keep the animal breathing in a range of 50-70 breaths per minute. The respiratory rate was continuously monitored with a respiration sensor placed under the animal's abdomen and connected to a computer running the small animal respiration-monitoring software. The mice temperature was maintained at 37°C with regulated water flow integrated to the stereotaxic bed and monitored with an endorectal probe.

**2.3 MRI hardware**

All mice were scanned with a 117/16 Biospec horizontal magnet (Bruker, Ettlingen, Germany) running Paravision 6.0.1 (Bruker, Ettlingen, Germany). The scanner was equipped with a BGA-9 gradient coil capable of reaching 752 mT/m on each axis and a quadrature surfacecCryoprobe™ (Bruker, Ettlingen, Germany) for both radiofrequency transmission and reception.

**2.4 MRI/MRS protocol**

In order to place the spectroscopic volume of interest (VOI), two-dimensional $T_2$-weighted sagittal and transversal images were acquired with a multi-slice multi-echo sequence (isotropic field of view, 16 mm; matrix size: 160 × 160; isotropic resolution, 0.1 mm; $T_R/T_E$, 1500/20 ms; total acquisition time, 4 min).

DW-MRS data were acquired in a VOI of 6 × 1.5 × 3 mm$^3$ including the body of the corpus callosum (Figure 2A), using a single-voxel stimulated echo (STE)-localization by adiabatic selective refocusing (LASER) sequence described previously (Ligneul et al., 2017). In this sequence, the diffusion module is disentangled from the localization module, thus yielding no cross-terms between diffusion and selection gradients.

The STE block consisted of three 90° pulses (pulse length of 100 μs), with de-phasing and re-phasing diffusion gradients located before the second 90° pulse and after the third 90° pulse, respectively. The LASER block consisted of a train of six slice-selective adiabatic 180° pulses (10-kHz pulse bandwidth). The sequence parameters were: $T_E/T_M$ = 33/60 ms, $T_R$ = 5 s, spectral width = 5 kHz and number of complex points = 4096. Diffusion-weighting was applied in three orthogonal directions ([1, 1, -0.5], [1, -0.5, 1], [-0.5, 1, 1] in the scanner coordinate system) with diffusion gradient duration = 3 ms, diffusion time $t_d$ = 64 ms and three increasing gradient strengths resulting in the *b*-values $b_1$ = 2, $b_2$ = 4 and $b_3$ = 6 ms/μm². A quasi-non-diffusion-weighted condition was also acquired with: $b_0$ = 0.03 ms/μm². Thirty-two averages were collected for each diffusion-weighted condition and saved as individual free induction decays for further processing. Water suppression was performed using variable power with optimized relaxation delays (VAPOR) and outer volume suppression (Tkac et al., 1999). For eddy current corrections, unsuppressed water reference scans were acquired using the same parameters as water suppressed spectra.

For spectral fitting, a macromolecule spectrum was acquired in a control mouse using a double inversion recovery ($T_{I1}$ = 2200 ms, $T_{I2}$ = 730 ms) STE-LASER sequence ($T_E$ = 33ms, $T_M$ = 50ms, $T_R$ = 4000ms), with *b*-value = 10 ms/μm².

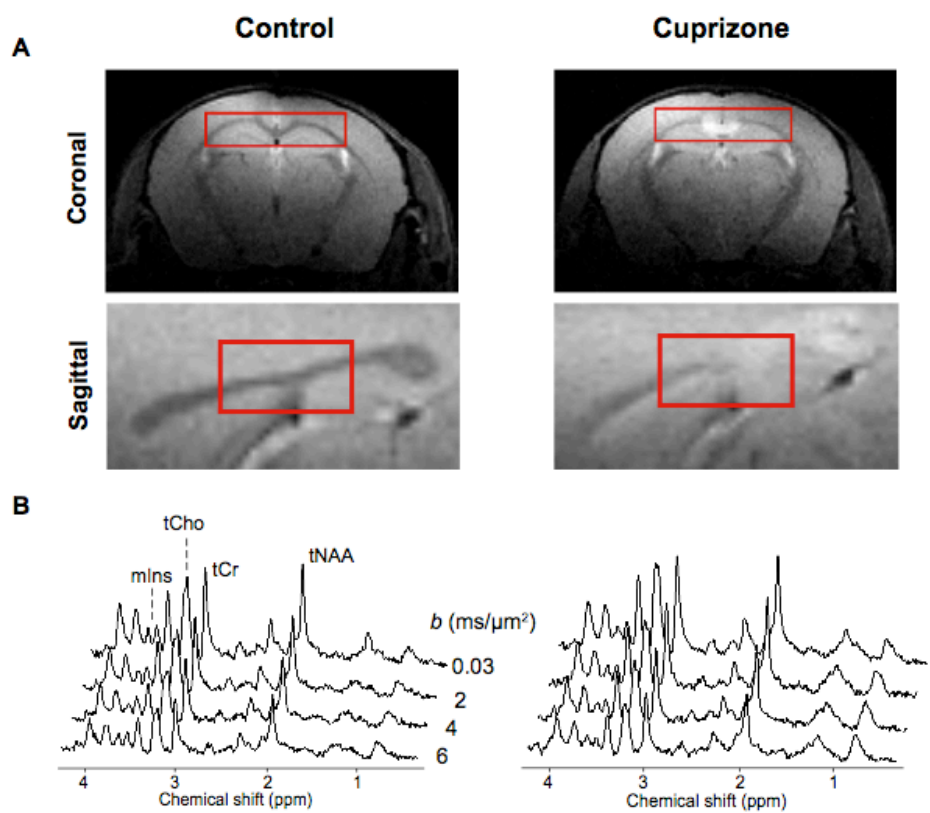

**Figure 2: Volumes of interest and DW-MRS spectra in control and CPZ mice.**

A) Location of the VOIs in the body of the corpus callosum shown on coronal and sagittal views of $T_2$-weighted images acquired in one control (left) and one CPZ (right) mice. B) DW-MRS spectra acquired up to $b = 6$ ms/µm$^2$ in direction [-0.5, 1, 1] in one control mouse (left) and one CPZ mouse (right).

mIns: myo-inositol; tCho: choline compounds ; tCr : creatine + phosphocreatine ; tNAA: *N*-acetylaspartate + *N*-acetylaspartylglutamate.

## 2.5 Spectral processing

All spectra were processed with an in-house written routine in MATLAB release R2016b (Mathworks, Natick, MA, USA). Single spectra were first corrected for eddy currents using water reference scans. Zero-order phase fluctuations and frequency drifts were corrected on single averages before summation using an area minimization and penalty algorithm and a cross-correlation algorithm, respectively (de Brouwer, 2009). For each *b*-value and diffusion-weighted direction, spectra were averaged across averages and subsequently analyzed using LCModel (Provencher, 1993) for metabolite quantification. The basis set included alanine, ascorbate, aspartate, Cr, γ-aminobutyric acid, glucose, glycine, glutamate (Glu), glutamine, glutathione, glycerophosphorylcholine (GPC), mIns, lactate, *N*-acetylaspartate, *N*-acetylaspartylglutamate, phosphocholine (PCho), PCr, phosphorylethanolamine, *scyllo*-inositol, taurine (Tau), threonine and experimentally measured macromolecules.

## 2.6 Metabolite diffusion and concentration measures

The ADCs of mIns, tCho, tCr and tNAA were calculated by linear regression of the logarithm of the signal decay in each diffusion direction. For the four metabolites, the average diffusivity was computed across the three diffusion-weighted directions.

tCr absolute concentrations were evaluated from the spectra measured at $b_0$, scaled using the water reference scan acquired at the same condition. Because tCr concentrations did not change significantly between CPZ and control mice (data not shown), the concentrations of mIns, tCho and tNAA were reported relative to the tCr concentrations, in order to reduce the variability due to partial volume effects and water relaxation rates. For completeness, ADCs and concentrations of Tau and Glu were also evaluated (Supplementary material).

**2.7 Immunohistochemistry**

Just after completion of the DW-MRS data collection, four CPZ and five control mice were sacrificed for IHC assessment. The animals were deeply anesthetized with a lethal dose of Euthasol (400 mg/ml, 100 mg/kg) prior to a transcardiac perfusion with 4% paraformaldehyde (PFA). The brains were rapidly removed and post-fixed in 4% paraformaldehyde for 24 hours, prior to cryoprotection in a 25% sucrose/phosphate buffer saline (PBS) solution. The brains were then frozen at -50°C using isopentane. Sagittal sections were cut using a cryostat with a thickness of 25 μm that were regularly spaced at 120-240 μm intervals.

Only one hemisphere was used and sections were collected from 0.96 mm relative to the Bregma until reaching the median line (-0.04 mm from the Bregma). Up to 40 tissue sections, of 25 μm each, per animal, were harvested. For analysis, 4 tissue sections at different levels from Bregma were selected such that they covered the same region of the brain for each animal.

**2.8 Immunohistochemistry data analysis**

From the IHC samples, we estimated the astrocyte (AAF), microglia (MAF) and oligodendrocyte area fractions (OAF) in the corpus callosum. For immunofluorescence analysis of astrogliosis, microgliosis and oligodendrocyte density the following antibodies were used respectively: anti-GFAP (polyclonal rabbit anti-glial fibrillary acidic protein, Agilent, 1:500 dilution), anti-IBA1 (rabbit anti-adaptive molecule linking ionizing calcium 1, Sobioda, 1:500 dilution) and anti-CC1 (adenomatous polyposis coli oligodendrocyte proteins, Anti-APC (Ab-7) mouse mAb (CC1), Merck, 1:100 dilution). A series consisting of ten slices were scanned with the X20 lens and Z-stacked with the Carl ZEISS Axio Scan.Z1 scanner. All analyses were performed in the body of the corpus callosum. The ImageJ software (Schneider et al., 2012) was used to quantify the cells of interest. Because the irregular shape of microglia and astrocytes did not allow the establishment of a count in two dimensions, a measurement of area seemed more suitable. The microglial and astrocyte area quantification was carried out by computing the ratio of the area of the stained region to the area of the total selected region.

**2.9 Statistical analysis**

Unpaired t-tests were performed to assess differences in metabolite ADCs, metabolite relative concentrations, and histology markers, between CPZ and control mice. Linear Pearson correlations between metabolite ADCs and AAF and MAF were evaluated both intra- and inter-cohort. A *p*-value < 0.05 was considered to represent statistically significant differences or statistically significant correlations. All statistical analyses were performed in MATLAB.

3. **Results**

**3.1 DW-MRS data quality**

Figure 2B shows the DW spectra acquired in the body of the corpus callosum in one control and one CPZ mouse at different *b*-values.

The Cramér-Rao lower bounds (CRLBs) associated with tNAA and tCr quantification were lower than 10% for all *b*-values. For mIns and tCho, the CRLBs were lower than 20% for all *b*-values, except for two control mice, where the CRLBs were higher than 30% for at least one *b*-value. MIns and tCho signals were therefore excluded from further analysis for these two mice. At $b_0$, the CRLBs were ~2% for tNAA and tCr, and ~6% for mIns and tCho. Figure 3 shows the signal logarithm attenuations for mIns, tCho, tCr and tNAA averaged across mice and directions, for both groups.

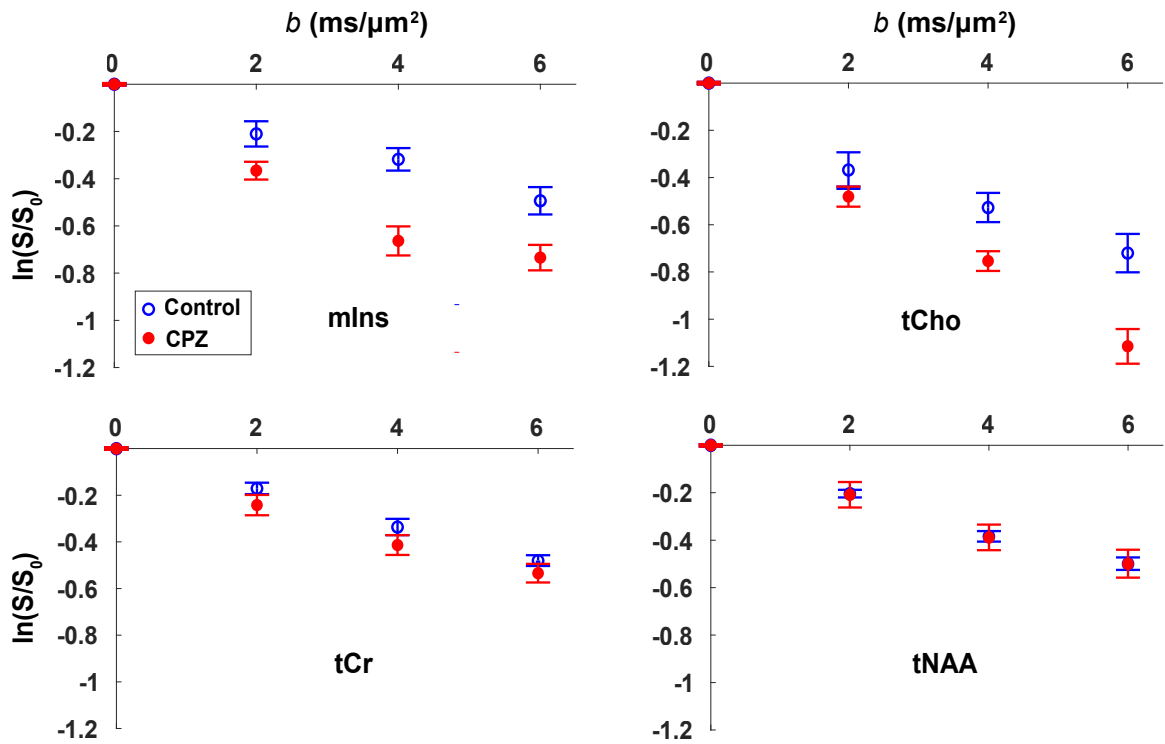

**Figure 3: Metabolite signal attenuation.** Average signal logarithm attenuation as a function of the *b*-value for mIns, tCho, tCr and tNAA in control (blue circles) and CPZ (red dots) mice. Data points and error bars stand for mean and standard error of the mean for each mouse group (n = 8 mice per group for mIns and tCho; n = 8 CPZ and 10 control mice for tCr and tNAA). Only mIns and tCho exhibit different diffusion behavior between the two mouse groups.

### 3.2 Apparent diffusion coefficients

Figure 4A shows metabolite ADCs evaluated for control and CPZ mice. The ADC of mIns was significantly higher in CPZ (0.130 ± 0.031 $\mu m^2/ms$, $p = 0.004$) with respect to control mice (0.082 ± 0.024 $\mu m^2/ms$). Similarly, the ADC of tCho was significantly higher in CPZ (0.188 ± 0.041 $\mu m^2/ms$, $p = 0.017$) with respect to control mice (0.120 ± 0.033 $\mu m^2/ms$). The ADCs of tNAA and tCr did not differ significantly between the two groups, neither did the ADCs of Tau and Glu (Inline Supplementary Figure S1A, Table S1). Metabolite ADCs are summarized in Table 1.

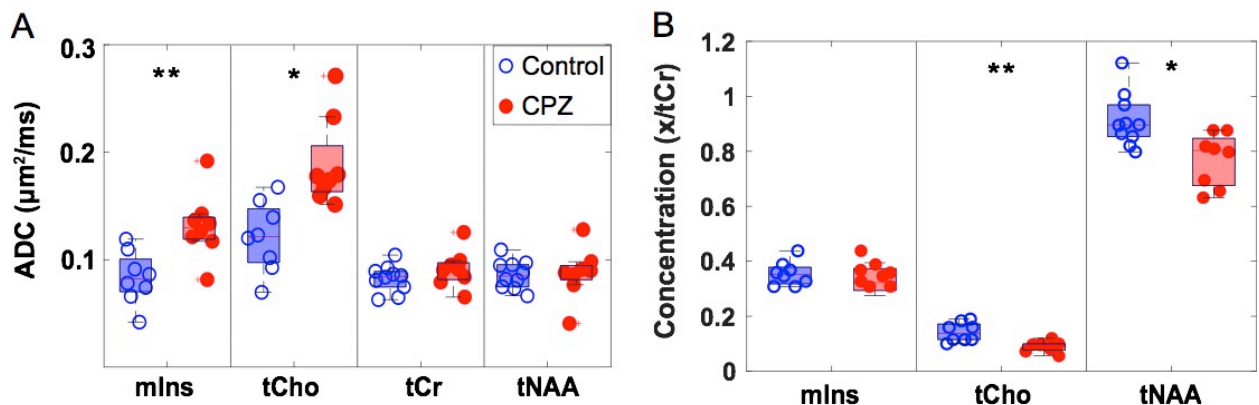

**Figure 4: Metabolite ADCs and concentrations in control and CPZ mice.** Box plots of metabolite ADCs (A) and concentrations relative to tCr (B) measured in control (blue circles) and CPZ (red dots) mice. For each box the central mark indicates the median ADC and the bottom and top edges indicate the 25th and 75th percentiles, respectively. Asterisks indicate significant difference between control and CPZ mice (*$p < 0.05$, **$p < 0.005$).

mIns: myo-inositol; tCho: choline compounds ; tCr : creatine + phosphocreatine ; tNAA: *N*-acetylaspartate + *N*-acetylaspartylglutamate.

### 3.3 Concentrations

Figure 4B shows metabolite concentrations relative to tCr, evaluated for control and CPZ mice. tNAA/tCr was significantly lower in CPZ ($0.770 \pm 0.096$, $p = 0.007$) with respect to control mice ($0.913 \pm 0.096$). tCho/tCr was also significantly lower in CPZ ($0.090 \pm 0.020$, $p = 0.002$) with respect to control mice ($0.142 \pm 0.035$). In contrast, the concentration of mIns did not differ significantly between groups (Table 1). In addition, Tau/tCr was significantly higher in CPZ ($1.108 \pm 0.095$, $p = 0.006$) with respect to control mice ($0.939 \pm 0.110$), while Glu/tCr was significantly lower in CPZ ($0.541 \pm 0.054$, $p = 0.003$) vs. control mice ($0.654 \pm 0.070$) (Inline Supplementary Figure S1B, Table S1).

**Table 1:** Mean and standard deviation of metabolite ADCs and concentrations relative to tCr evaluated for each mouse group. The *p*-values are reported in the last column.

|  | Metabolite | CPZ (n = 8) | Control (n = 10) | *p*-value |
|---|---|---|---|---|
| **ADC ($\mu m^2$/ms)** | tNAA | 0.086 ± 0.023 | 0.085 ± 0.014 | 0.898 |
|  | tCr | 0.089 ± 0.016 | 0.081 ± 0.013 | 0.238 |
|  |  | **CPZ (n = 8)** | **Control (n = 8)** |  |
|  | tCho | 0.188 ± 0.041 | 0.120 ± 0.033 | **0.017** |
|  | mIns | 0.130 ± 0.031 | 0.082 ± 0.024 | **0.004** |
| **Concentration** |  | **CPZ (n = 8)** | **Control (n = 10)** |  |
|  | tNAA/tCr | 0.770 ± 0.096 | 0.913 ± 0.096 | **0.007** |
|  |  | **CPZ (n = 8)** | **Control (n = 8)** |  |
|  | tCho/tCr | 0.090 ± 0.020 | 0.142 ± 0.035 | **0.002** |
|  | mIns/tCr | 0.331 ± 0.045 | 0.355 ± 0.044 | 0.294 |

Bold *p*-values indicate statistically significant differences.

mIns: myo-inositol; tCho: choline compounds ; tCr : creatine + phosphocreatine ; tNAA: *N*-acetylaspartate + *N*-acetylaspartylglutamate.

### 3.4 Correlations with immunohistochemistry markers

Representative IHC images for astrocytic (GFAP) and microglial (IBA1) immunodetection in the corpus callosum of control and CPZ mice are shown in Figure 5A. AAF and MAF quantified in the body of the corpus callosum are reported for CPZ and control mice in Figure 5B. Significantly higher AAF and MAF were observed in CPZ (22.7 ± 3.8 % and 24.3 ± 5.7 % respectively, *p* < 0.001 for both) with respect to control mice (7.5 ± 1.6 % and 7.2 ± 1.5 % respectively). Also, a significant lower OAF was observed in CPZ (15.7 ± 2.4 %, *p* = 0.003) with respect to control mice (25.3 ± 3.7 %) (Inline Supplementary Figure S2).

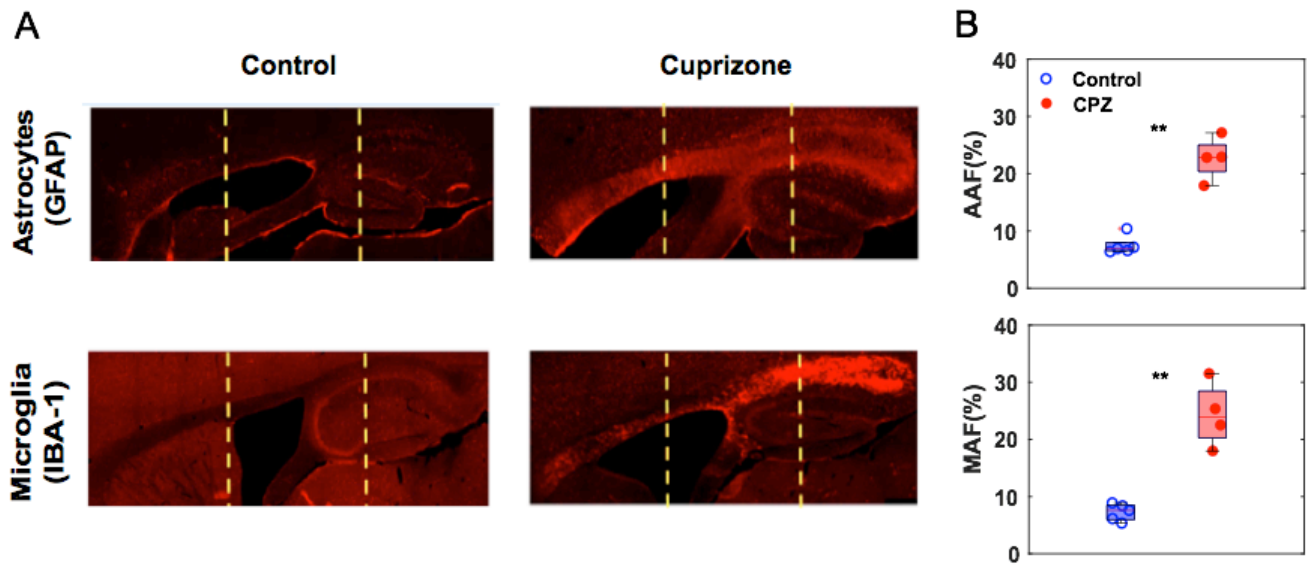

**Figure 5: Immunohistochemistry markers of glial cells in control and CPZ mice.** A) Immunohistochemistry images for astrocytic (GFAP) and microglial (IBA1) immunofluorescence obtained in the corpus callosum of one control mouse (left) and one CPZ mouse (right). Yellow dashed lines delimit the body of the corpus callosum, where AAF and MAF were calculated. B) Box plots of AAF and MAF measured in the body of the corpus callosum of control (blue circles) and CPZ (red dots) mice. For each box the central mark indicates the median value and the bottom and top edges indicate the 25th and 75th percentiles, respectively. Asterisks indicate significant difference between control and CPZ mice (**$p$ < 0.005). AAF: astrocyte area fraction; MAF: microglia area fraction; GFAP: glial fibrillary acidic protein; IBA1: adaptive molecule linking ionizing calcium 1.

Figure 6 shows correlations between the IHC glial markers and the ADCs of mIns (6A, C) and tCho (6B, D). As expected, the CPZ data showed a larger spread with respect to the control data, likely reflecting variable inflammatory response in CPZ mice. Significant correlations were observed between the ADC of mIns and both AAF (Pearson correlation coefficient ρ = 0.84, *p* = 0.004; Figure 6A) and MAF (Pearson correlation coefficient ρ = 0.80, *p* = 0.009; Figure 6C). Notably, a stronger linear correlation was observed between the ADC of mIns and AAF (Pearson correlation coefficient

ρ = 0.99, *p* = 0.009) when considering the CPZ cohort alone (Figure 6A). AAF and MAF were also significantly correlated (Pearson correlation coefficient ρ = 0.91, *p* < 0.001, Inline Supplementary Figure S3), indicating a strong interplay between astrocytic and microglia activation.

The correlations between the ADC of tCho and both AAF (Pearson correlation coefficient ρ = 0.62, *p* = 0.077; Figure 6B) and MAF (Pearson correlation coefficient ρ = 0.61, *p* = 0.082; Figure 6D), were just above the threshold for statistical significance. No significant correlations were observed between the ADC of tCho and neither of the IHC glial markers when considering the CPZ group alone.

No statistically significant correlations were observed between the IHC glial markers and tCr nor tNAA ADCs (data not shown).

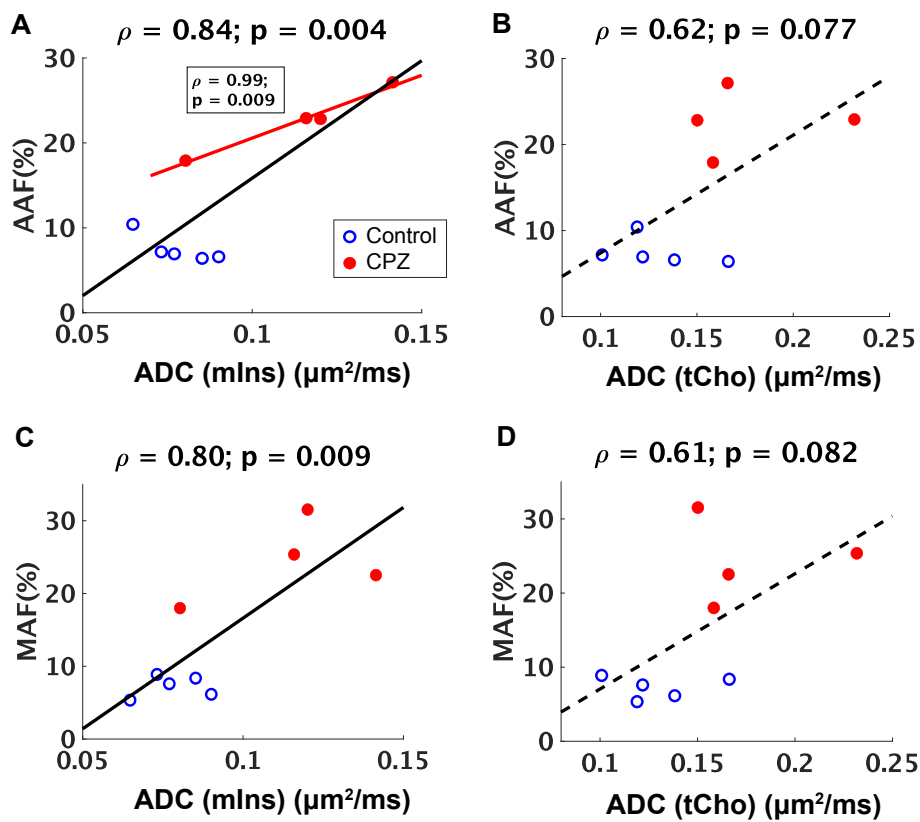

**Figure 6: Correlations between DW-MRS and immunohistochemistry markers of glial cells**. Scatter plots of AAF and MAF against mIns ADC (A, C) and tCho ADC (B, D). Only half of the mice were sacrificed for IHC analysis. Black lines and dashed lines indicate significant linear

regressions (*p* < 0.05) and trends (*p* < 0.10), respectively. Corresponding Pearson correlation coefficients and *p*-values are reported on top of each plot. In (A), the red line indicates the linear regression calculated considering the CPZ group alone. The corresponding Pearson correlation coefficient and *p*-value are reported in the box.

AAF: astrocyte area fraction; MAF: microglia area fraction; mIns: myo-inositol; tCho: choline compounds ; tCr : creatine + phosphocreatine ; tNAA: *N*-acetylaspartate + *N*-acetylaspartylglutamate.

## 4. Discussion

**4.1 mIns and tCho diffusion probe ongoing inflammation processes in CPZ mice**

We reported the first investigation of white matter glial cell reactivity in the CPZ mouse model using in vivo DW-MRS. The main finding of our study was the observation of significant alterations in the diffusivity of both glial metabolites under investigation, mIns and tCho, measured in the corpus callosum of CPZ-fed mice after 6 weeks of intoxication, and compared to control mice. The ADCs of mIns and tCho were significantly higher in the CPZ group, suggesting that these molecules were diffusing in larger intra-cellular compartments. Because mIns and tCho are known to be located mainly in glial cells (Choi et al., 2007; Urenjak et al., 1993), this finding was consistent with the presence of hypertrophic glial cells, hallmarks of ongoing inflammation, in the corpus callosum of CPZ mice. Despite the complexity of metabolic dynamics in the intra-cellular environment, previous studies have shown evidence that the fraction of metabolites confined in sub-cellular compartments is negligible with respect to metabolites diffusing in the cellular cytoplasm (Ligneul et al., 2017; Najac et al., 2016; Palombo et al., 2017, 2016; Valette et al., 2018). Thus, the metabolite displacement probed by DW-MRS mostly reflects the microstructure of the cells in which metabolites are compartmentalized. Under our experimental conditions, (e.g. maximum $b$-value = 6 ms/μm$^2$, $t_d$ = 64 ms), it was not possible to extract from the DW-MRS data quantitative morphological parameters such as soma radii, fiber diameters and length, however, variations in metabolite diffusion allowed to capture qualitative alterations in cell size.

Our in vivo findings were corroborated by IHC data, revealing significantly higher AAF and MAF in the CPZ vs. the control mouse group. In addition, mIns ADC showed a significant positive correlation with both AAF and MAF histological markers of glial cells, when the two mouse groups were pooled together. Remarkably, when considering only the CPZ mouse group, the ADC of mIns showed even a more striking positive correlation with AAF (Figure 6A). This result was in line with a recent study suggesting mIns diffusion as a specific intra-cellular marker of astrocytic hypertrophy in a mouse model of "pure" astrocyte reactivity induced by the cytokine ciliary neurotrophic factor (CNTF) (Ligneul et al., 2019). Because in our study AAF and MAF were also significantly correlated (Inline Supplementary Figure S3), the significant correlation between mIns ADC and MAF was not inconsistent with the expected specificity of mIns to astrocytes.

In the CNTF model, where the presence of hypertrophic astrocytes is not associated with neuronal death or with microglial activation (Escartin, 2006), Ligneul et al. (2019) reported elevated mIns diffusion and no significant changes in tCho diffusivity compared to control mice. In view of this previous report, and given our observation of a trend to a positive correlation between tCho ADC and MAF (Figure 6D), it is tempting to speculate that the elevated tCho diffusivity observed in our study in the CPZ mouse model compared to control mice may be specifically linked to microglia reactivity. Remarkably, our findings were very consistent with the human data obtained in patients with neuropsychiatric systemic lupus erythematosus (NPSLE), an inflammatory autoimmune disease characterized by morphological changes in microglia and astrocytes (Ercan et al., 2016). The differences in tCho ADC between diseased and control groups were even greater in our study, in line with the fact that CPZ intoxication likely induced wider and more pronounced inflammation compared to that observed in NPSLE patients. Nevertheless, we did not observe a significant correlation between tCho ADC and MAF when considering the CPZ group alone, possibly due to the small number of CPZ mice examined. The hypothesis that tCho diffusion is a specific marker of microglia reactivity needs therefore to be corroborated by further experiments. Additionally, the tCho signal is generated by the contributions of GPC and PCho (while the contribution from free Cho is negligible in the brain). As a consequence, a change in the ratio between GPC and PCho concentrations, possibly associated with altered phospholipid metabolism in CPZ mice, would also yield a variation in tCho diffusion, due to the intrinsic different diffusivities of the GPC and PCho molecules.

Notably, the unchanged diffusivity of the intra-axonal metabolite tNAA in CPZ vs. control mice, suggested that the intra-axonal compartments were still intact after 6 weeks of CPZ diet and under our experimental conditions, while glial reactivity was the dominant ongoing process in response to CPZ intoxication at this time point. This result was consistent with the presence of partial demyelination associated with still intact axonal diameters quantified using EM in the same mouse group and reported recently in another study (Hill et al., 2019).

The ADC of tCr was also not significantly different between CPZ and healthy mice. Because tCr is located both in glia and neurons, this negative finding may reflect the concomitant presence of competing mechanisms that would independently lead to higher or lower tCr ADC values. In fact, higher tCr ADCs have been previously attributed to glial reactivity in patients with NPSLE (Ercan et al., 2016), while lower tCr ADC, possibly due to an increased PCr/Cr ratio, have been linked to energy dysregulation processes in the normal appearing white matter of patients with multiple sclerosis (MS) compared to healthy subjects (Bodini et al., 2018). Mitochondrial impairment has been suggested to significantly contribute to oligodendroglial apoptosis in CPZ-treated mice, thus compromising energy metabolism of cells (Pasquini et al., 2007). The combination of severe inflammation with impairment of energy metabolism in CPZ mice may prevent the use of tCr diffusion as a suitable marker of cell damage in this mouse model.

**4.2 Metabolite concentrations are not specific markers of inflammation in CPZ mice**

We measured significantly lower tNAA/tCr, tCho/tCr and Glu/tCr and significantly higher Tau/tCr in CPZ vs. control mice, while no significant differences were observed in mIns/tCr concentrations between groups. These results were in excellent agreement with a previous MRS study conducted under similar experimental conditions as in our study (Orije et al., 2015). Lower tNAA/tCr and Glu/tCr levels may be attributed to reversible axonal suffering caused by demyelination or mitochondrial dysfunction induced by the CPZ diet. In contrast, the results obtained for the glial metabolite concentrations are difficult to interpret and cannot not be directly linked to ongoing inflammation. Although inflammation is known to be associated with glial cell proliferation, which should potentially induce an increase of both tCho and mIns levels, previous MRS studies conducted in patients with MS are controversial, showing elevated, reduced or unchanged tCho levels (Gustafsson et al., 2007; Oh et al., 2004; Sajja et al., 2009). The CPZ model is characterized by a failure in the energy metabolism due to mitochondrial dysfunction (Kipp et al., 2009), which could be the cause of tNAA, Glu and tCho reduced levels compared to control mice (Bianchi et al., 2003). Instead, the lack of significant differences in mIns levels between the two mouse groups was somehow surprising given that astrogliosis is typically associated with elevated mIns concentrations. The unchanged mIns concentration in CPZ vs. control mice may indicate that, after 6 weeks of CPZ intoxication, glial cell hypertrophy and proliferation could have been accompanied by a concomitant decrease in local mIns levels, which contributed to maintain apparent stable concentrations of this metabolite. This finding suggested that mIns concentration was not a suitable marker of glial reactivity, at least in this mouse model.

Finally, the severe loss of oligodendroglial cells, corroborated by the observed lower CC1 immunodetection in CPZ vs. control mice (Inline Supplementary Figure S2), likely contributed to the lower or unchanged concentrations of both the glial metabolites and tNAA (which is known to be present at low concentrations also in oligodendrocytes).

### 4.3 Limitations and further perspectives

The major limitation of this study was that the AAF and MAF histological metrics evaluated the total area fractions occupied by astrocytes and microglia, respectively, and did not enable a distinction between cell size increase due to hypertrophy and increase in cell number due to cell proliferation. Because metabolite diffusion is known to reflect changes in cell morphology rather than proliferation, this may have slightly impacted the correlations between metabolite diffusion measures and histological metrics.

The second major limitation of this study relied on the fact that the spectroscopic voxel included a non-negligible content of the grey matter surrounding the corpus callosum, possibly leading to partial volume effects. This could not be avoided as a smaller voxel size covering mostly the corpus callosum would have not enabled sufficient signal-to-noise ratio and reliable metabolite quantification at high *b*-values. Although it is known that the corpus callosum is the most affected brain area in the CPZ model, we cannot not exclude that inflammation may have extended to grey matter areas. If this was the case, we would expect a similar reactivity of glial cells as for the white matter, which may have further contributed to the observed abnormally high mIns and tCho ADCs. Further studies will include histological assessment of glial alterations also in the grey matter regions included in the spectroscopic voxel, as well as quantification of the glial cell number for proper estimation of both cell size and density variations. In addition, to further validate the role of mIns and tCho diffusion metrics as reliable markers of inflammation, future studies shall focus on the longitudinal variations of DW-MRS measures in the CPZ mouse model during different stages of de- and re-myelination.

In addition, in order to test the hypothesis of tCho diffusion as specific marker of microglia activation, loss-of-function experiments in which astrocytes or microglia are selectively depleted in mice treated with CPZ are warranted.

## 5. Conclusions

This study demonstrated that mIns diffusion could be used as a powerful marker of inflammation-driven astrocytic hypertrophy in CPZ mice and suggested that tCho diffusion may be a possible specific probe of microglia reactivity. While mIns diffusion data showed an excellent correlation with histologic measures of astrocytic activation even in a very small mouse group, the validation of tCho diffusion data with histological markers of microglia should be corroborated in a larger cohort and potentially in a more specific mouse model of microglia reactivity. This work was conducted under experimental conditions that could be translated into the clinical setting and poses the bases for the study of inflammation processes in the human brain. In perspective, measuring longitudinal changes of DW-MRS markers of inflammation may be extremely useful for assessing disease progression, evaluating optimal therapeutic temporal windows, and monitoring the effect of specific treatments targeting inflammation.

## 6. Acknowledgments

The authors acknowledge the support from the programs 'Institut des neurosciences translationnelles' [ANR-10-IAIHU-06] and 'Infrastructure d'avenir en Biologie Santé' [ANR-11-INBS-0006].
MP is supported by EPSRC [EP/N018702-1] and UKRI Future Leaders Fellowship [MR/T020296/1].
JV acknowledges funding from the European Research Council (ERC) under the European Union's FP7 and Horizon 2020 research and innovation programmes [grant agreements No 336331 and 818266].

# Supplementary material

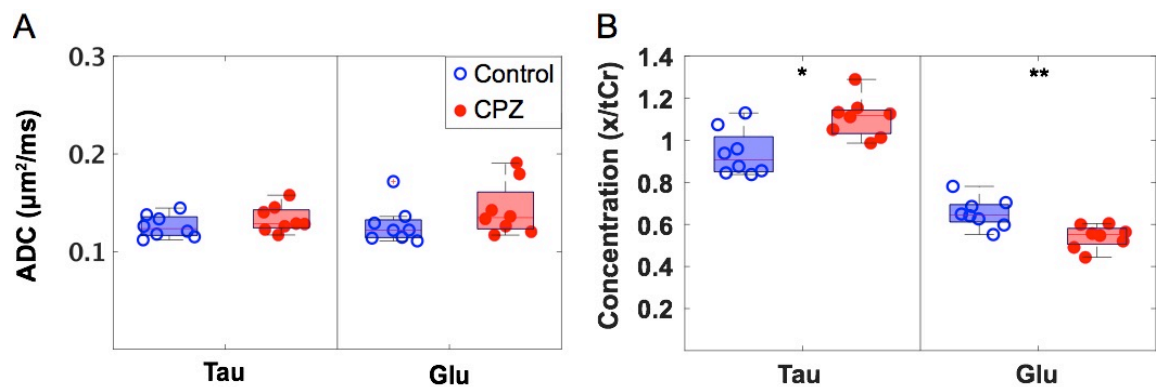

**Figure S1: Taurine and glutamate ADCs and concentrations in control and CPZ mice.** Box plots of taurine and glutamate ADCs (A) and concentrations relative to tCr (B) measured in control (blue circles) and CPZ (red dots) mice. For each box the central mark indicates the median ADC and the bottom and top edges indicate the 25th and 75th percentiles, respectively. Asterisks indicate significant difference between control and CPZ mice (*p < 0.05, **p < 0.005). Tau: taurine; Glu: glutamate.

**Table 1S:** Mean and standard deviation of taurine and glutamate ADCs and concentrations relative to tCr evaluated for each mouse group. The *p*-values are reported in the last column.

|  | Metabolite | CPZ (n = 8) | Control (n = 8) | *p*-value |
|---|---|---|---|---|
| ADC ($\mu m^2/ms$) | Tau | 0.133 ± 0.013 | 0.126 ± 0.012 | 0.268 |
|  | Glu | 0.143 ± 0.027 | 0.127 ± 0.020 | 0.210 |
| Concentration | Tau/tCr | 1.108 ± 0.095 | 0.939 ± 0.110 | **0.006** |
|  | Glu/tCr | 0.541 ± 0.054 | 0.654 ± 0.070 | **0.003** |

Bold *p*-values indicate statistically significant differences.

Tau: taurine; Glu: glutamate.

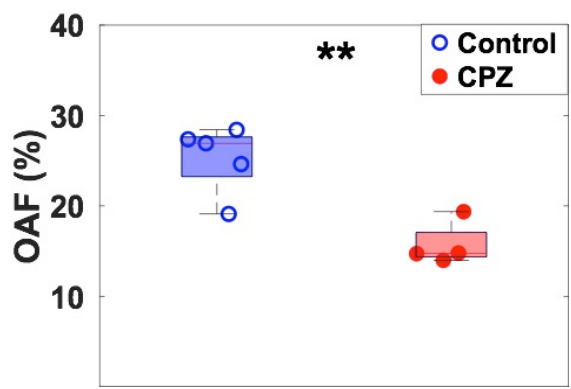

**Figure S2: Immunohistochemistry oligodendrocyte markers in control and CPZ mice.**

Box plots of OAF measured in the body of the corpus callosum of control (blue circles) and CPZ (red dots) mice. For each box the central mark indicates the median value and the bottom and top edges indicate the 25th and 75th percentiles, respectively. Asterisks indicate significant difference between control and CPZ mice (**$p < 0.005$). OAF: oligodendrocyte area fraction.

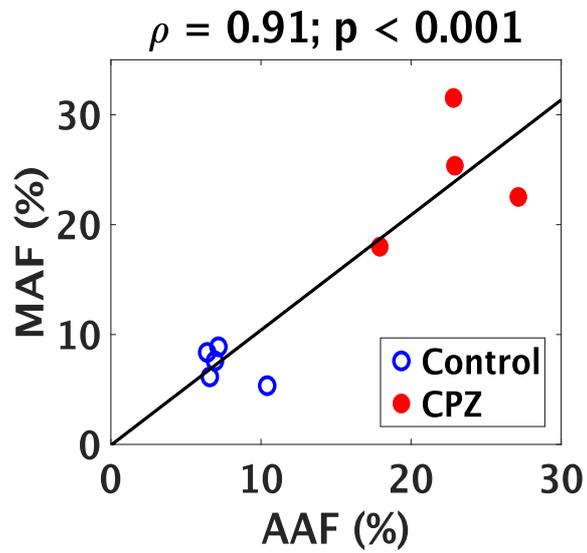

**Figure S3: Correlation between astrocytic and microglial immunohistochemistry markers.** Scatter plots of MAF against AAF. Only half of the mice were sacrificed for IHC analysis. The black line indicates a significant linear regression. The corresponding Pearson correlation coefficient and *p*-value are reported on top. AAF: astrocyte area fraction; MAF: microglia area fraction.